\documentclass[onecolumn,showpacs,amsmath,amssymb,pre]{revtex4-1}

\usepackage{graphicx}
\usepackage{color}
\usepackage{dcolumn}
\usepackage{amsmath}

\usepackage{sidecap}
\usepackage{dsfont}

\usepackage{mathrsfs}
\usepackage{amsfonts}
\usepackage{indentfirst}
\usepackage{bm}
\usepackage{mathtools}

\usepackage[colorlinks,citecolor=blue]{hyperref}

\newcommand{\be}{\begin{equation}}
\newcommand{\ee}{\end{equation}}
\newcommand{\bey}{\begin{eqnarray}}
\newcommand{\eey}{\end{eqnarray}}
\newcommand{\bw}{\begin{widetext}}
\newcommand{\ew}{\end{widetext}}

\newcommand{\ra}{\rangle}
\newcommand{\la}{\langle}

\newcommand{\ba}{\begin{array}}
\newcommand{\ea}{\end{array}}
\newcommand{\bi}{\begin{itemize}}
\newcommand{\ei}{\end{itemize}}
\newcommand{\bem}{\begin{enumerate}}
\newcommand{\eem}{\end{enumerate}}

\begin{document}

\title{Multifractality in quasienergy space of coherent states 
as a signature of quantum chaos}

\author{Qian Wang $^{2,1}$ and Marko Robnik $^{1}$}

\address{
$^{1}$ \quad CAMTP-Center for Applied Mathematics and Theoretical Physics, 
University of Maribor, SI-2000 Maribor, Slovenia \\
$^{2}$ \quad Department of Physics, Zhejiang Normal University, Jinhua 321004, China}

\begin{abstract}
We present the multifractal analysis of coherent states in kicked top model 
by expanding them in the basis of Floquet operator eigenstates. 
We demonstrate the manifestation of phase space structures in the 
multifractal properties of coherent states.
In the classical limit, the classical dynamical map can be constructed, 
allowing us to explore the corresponding
phase space portraits and to calculate Lyapunov exponent. 
By tuning the kicking strength, the system undergoes a transition from regularity to chaos. 
We show that the variation of multifractal dimensions of coherent states with kicking strength is able to 
capture the structural changes of the phase space.   
The onset of chaos is clearly identified by the phase space averaged multifractal dimensions, which are well described by
random matrix theory in strongly chaotic regime. 
We further investigate the probability distribution of expansion coefficients, and show that the deviation between
the numerical results and the prediction of random matrix theory behaves as a reliable detector of quantum chaos.

\end{abstract}

\date{\today}
 
\maketitle

\section{Introduction}

Quantum chaos plays a crucial role
in many fields of physics, such as quantum statistics 
\cite{Altland2012,Nandkishore2015,LucaD2016,Borgonovi2016,Deutsch2018}, 
quantum information science 
\cite{Vidmar2017,Alicki1996,WangX2004,Ghose2008,Piga2019,Lerose2020,Lantagne2020,Anand2021}, 
and high energy physics \cite{Maldacena2016,Magan2018,Jahnke2019}.
In particular, chaos of interacting quantum systems, dubbed as many-body quantum chaos, 
has attracted significant attention in recent years 
\cite{Stanford2016,ChanA2018,Kos2018,Bertini2018,Friedman2019,Kobrin2021,Bertini2019}. 
However, in contrast to the classical chaos, which is well defined as the hypersensitivity to the initial condition
\cite{Cvitanovic2005,Lichtenberg2013,Arnold2013},
the definition of the quantum chaos in time-dependent domain is still lacking, due to the fact that there is 
no quantum analog of classical trajectories in general quantum theory.
{In this regard, studies of OTOC (out-of-time ordered correlator) are highly relevant (see Sec. \ref{Third}).}
Therefore, the questions of how the chaotic dynamics manifests itself in quantum systems 
and how to diagnose the quantum chaos immediately and naturally arise.

There are several ways to detect quantum chaos, which probe the effects of chaos on quantum systems from
different aspects, the most popular one being the level spacing statistics 
\cite{Berry1977,BGS1984,Stockmann2000,Brody1981,Haake2010,Izrailev1990,Oganesyan2007,Atas2013}. 
The BGS conjecture \cite{BGS1984} allows us to identify a given quantum system as chaotic system
when its level spacing statistics is identical to the prediction of random matrix theory (RMT) \cite{Mehta2004}. 
Besides the level spacing statistics, the statistics of eigenvectors of quantum Hamiltonian 
can also be used as a benchmark to certify quantum chaos
\cite{Izrailev1990,Berry1977a,Porter1956,Kus1988,Alhassid1986,Alhassid1989,HaakeZ1990}. 
For quantum chaotic systems, their eigenfunction statistics is also well described by RMT. 

A drawback of the above mentioned quantum chaos indicators is that
they only unveil the overall behaviors and cannot probe local properties of quantum chaotic systems.
Since a generic system usually has structured phase space with coexistence of regular and chaotic regions
rather than featureless fully developed chaotic region, 
it is therefore highly desirable to investigate such quantities
which enable us to analyze the local chaotic behaviors of a quantum system. 
With the help of coherent states (or localized wave-packets), the local chaotic behaviors of quantum systems 
have been extensively explored in a variety of works
\cite{Zyczkowski1990,Leboeuf1990a,Batistic2013,Magnani2016,Magnani2017,Mondal2020,QianWg2020,Villasenor2021}. 
Here, by considering the kicked top model, we are interested in how to reveal the phase space structures and 
the degree of chaos by means of multifractality of coherent states. 

As a general phenomenon in nature, multifractality characterizes a wide range of complex phenomena from
turbulence \cite{Sreenivasan1991} to the chemistry \cite{Stanley1988} and financial markets \cite{Jiang2019}. 
It has been proven that the multifractal analysis also acts as a powerful tool to understand disorder induced 
metal-insulator transition in both single- and many-particle Hamiltonians 
\cite{Mirlin2000,Tarzia2020,Ever2008,Rodriguez2011,Mace2019,Solorzano2021}.  
The multifractality is also presented in the ground state of quantum many-body system and determines the
physics of ground state quantum phase transition \cite{AtasY2013,Luitz2014,Lindinger2019,AtasY2014}.
In addition, multifractal analysis of quantum states of random matrix models
\cite{Martin2010,Bogomolny2011,Graca2012,Tomasi2020,Bilen2021}, 
chaotic quantum many-body systems \cite{Backer2019,Pausch2021} 
and open quantum systems \cite{Bilen2019} were studied before.
In the present work, the fractal properties of the coherent states are examined in order to
identify both the local and global signatures of quantum chaos.

We perform multifractal analysis of coherent states by expanding them in the basis of the 
eigenstates of the Floquet operator.
To quantify the character of multifractality, we consider the so-called multifractal dimensions $D_q$,
which characterize the structure of a quantum state in Hilbert space.
For fully chaotic states $D_q=1$, for localized states $D_q=0$ with $q\geq 0$, while for the 
multifractal states $0<D_q<1$ is a function of $q$ \cite{Solorzano2021,Pausch2021}.
In the kicked top model, we show that the multifractal properties of coherent states 
strongly depend on the chaotic behavior of its classical counterpart.
We find that the multifractal dimensions exhibit a similar transition 
as observed in phase space portraits and Lyapunov exponents when the system varies from regular to 
mixed phase phase and globally chaotic dynamics.
In particular, we demonstrate that the structure of classical mixed phase space 
can be clearly distinguished by the properties of multifractal dimensions.
We also show that coherent states within the strong chaotic regime become ergodic 
as the system size goes to infinity, as expected from RMT predictions. 
On the contrary, coherent states in regular regime are still behaved 
as the multifractal states even in the thermodynamic limit. 
By exploring the probability distribution of the expansion coefficients, we demonstrate why the multifratal dimensions
of coherent states are not zero in the regular regime and why RMT predictions on the behavior of multifractal dimensions   
are reliable in the fully chaotic regime.

The remainder of this article is organized as follows. 
In Sec.~\ref{Secd}, we introduce the kicked top model, derive the stroboscopic evolution 
of the angular momentum for both quantum and classical cases, and analyze 
the classical and quantum chaotic behaviors as well. 
In Sec.~\ref{Third}, we present our numerical results in detail for the multifractal analysis of coherent states,
discuss the manifestation of phase space features and onset of chaos in
behavior of multifractal dimensions. 
Finally, we make some concluding remarks and summarize our results in Sec.~\ref{Forth}.

 \section{Kicked top model} \label{Secd}

 As a paradigmatic model for both theoretical 
 \cite{Haake1987,Fox1994,Alicki1996,WangX2004,Ghose2008,Trail2008,Lomardi2011,Piga2019,Sieberer2019,
 Dogra2019,Lerose2020,Arias2020,Arias2021} 
 and experimental \cite{Chaudhury2009,Neill2016,Tomkovic2017,Meier2019} 
 studies of quantum chaos,
 the kicked top model consists of a larger spin with total angular momentum $j$
 whose dynamics is captured by the following Hamiltonian (throughout this work, $\hbar=1$)
 \cite{Haake1987,Piga2019}
 \be \label{KTH}
    H=\alpha J_x+\frac{\kappa}{2j}J_z^2\sum_{n=-\infty}^{n=+\infty}\delta(t-n),
 \ee
 where $J_{a} (a=x,y,z)$ are the components of the angular momentum operator ${\bf J}$.
 The first term in the Hamiltonian represents the free precession of the spin around the $x$ axis at a rate $\alpha$,
 while the periodic $\delta$ kicks with strength $\kappa$, the second term in Eq.~(\ref{KTH}), 
 periodically generates an impulsive rotation about the $z$ axis by an angle
 $(\kappa/2j)J^2_z$, with $n$ being the number of kicks. Here, the time period between two successive kicks 
 has been set to unity.
 The time evolution operator corresponding to above Hamiltonian is
 the Floquet operator \cite{Haake1987}
 \be \label{Floquet}
   \mathcal{F}=\exp\left(-i\frac{\kappa}{2j}J_z^2\right)\exp(-i\alpha J_x).
 \ee
 
 In the numerical calculation, the Floquet operator should be expressed in a certain representation.
 To this end, we employ the Dicke states $\{|j,m\rangle; (m=-j,-j+1,\ldots,j)\}$, that satisfy
 $\mathbf{J}^2|j,m\ra=j(j+1)|j,m\ra$ and $j_z|j,m\ra=m|j,m\ra$.
 Then, the matrix elements of the Floquet operator are given by
 \be
    \la j,m|\mathcal{F}|j,m'\ra=\exp\left[-i\frac{\kappa}{2j}m^2\right]d_{mm'}^{(j)}(\alpha),
 \ee
 where
 \be
    d_{mm'}^{(j)}=\la j,m|e^{-i\alpha J_x}|j,m'\ra=\sum_{k_x=-j}^{k_x=j}e^{-i\alpha k_x}\la j,m|j,k_x\ra\la j,k_x|j,m'\ra,
 \ee
 is the so-called Winger $d$-function \cite{Fox1994} with
 $|j,k_x\ra$ being the eigenstates of $J_x$, so that $J_x|j,k_x\ra=k_x|j,k_x\ra$ and $-j\leq k_x\leq j$.
 As the magnitude of spin operator is a conserved quantity so that the 
 matrix dimension is equal to $2j+1$.  
 Moreover, as the Floquet operator in Eq.~(\ref{Floquet}) also conserves parity $\Pi=e^{i\pi(J_x+j)}$, 
 its matrix space can be further split into 
 even- and odd-parity subspaces with dimensions $\mathcal{D}_{even}=j+1$ and $\mathcal{D}_{odd}=j$, respectively.

 For an arbitrary initial state $|\psi_0\rangle$, the evolved state after $n$th kick is given by
 \be
       |\psi_n\rangle=\mathcal{F}^n|\psi_0\rangle.
 \ee
 The expectation values of the angular momentum operators are, therefore, evolved as follows
 \be
    \la J_a(n)\ra=\la\psi_n|J_a|\psi_n\ra=\la\psi_0|\mathcal{F}^{\dag, n} J_a(0)\mathcal{F}^n|\psi_0\ra,
 \ee
 where $J_a(n) (a=x,y,z)$ denotes the $a$th components of the spin operator $\mathbf{J}$ at $t=n$. 
 Hence, the stroboscopic evolution of the spin operators can be written as
 \be 
     J_a(n+1)=\mathcal{F}^\dag J_a(n)\mathcal{F}.
 \ee
 By using the operator identity,
 \be
     e^{\lambda{A}}{B}e^{-\lambda{A}}={B}+\lambda[A, B]+\frac{\lambda^2}{2}[A,[A,B]]+\ldots
 \ee
 the explicit form of the quantum iterated map reads \cite{Haake1987,Fox1994,Arias2021}
 \begin{align}
    J_x(n+1)=&\frac{1}{2}\{J_x(n)+i[J_y(n)\cos\alpha-J_z(n)\sin\alpha)\} \notag \\
              &\times\exp\left[i\frac{\kappa}{2j}\left\{2[J_y(n)\sin\alpha+
       J_z(n)\cos\alpha]+1\right\}\right]+\mathrm{H.c.} \\
    J_y(n+1)=&\frac{1}{2i}\{J_x(n)+i[J_y(n)\cos\alpha-J_z(n)\sin\alpha)\} \notag \\
              &\times\exp\left[i\frac{\kappa}{2j}\left\{2[J_y(n)\sin\alpha+
       J_z(n)\cos\alpha]+1\right\}\right]+\mathrm{H.c.} \\
    J_z(n+1)=&J_y(n)\sin\alpha+J_z\cos\alpha.   
 \end{align}

 \subsection{Classical kicked top} \label{SubA}
 
 The classical counterpart of the kicked top model can be obtained in the limit $j\to\infty$. 
 To show this, we first introduce the scaled spin operators $S_a=J_a/j$, 
 which behave as the classical variables due to the vanishing commutators between them as $j\to\infty$. 
 Then, by factorizing the mean values of the products of the angular momentum operators as 
 $\la J_aJ_b\ra/j^2=S_aS_b$ \cite{Ghose2008,Arias2021,WangXq2010},
 it is straightforward to find that the stroboscopic map of the classical angular momentum  
 can be written as \cite{Piga2019}
 \be \label{CMP}
   \begin{bmatrix}
     S_x(n+1) \\
     S_y(n+1) \\
     S_z(n+1)
   \end{bmatrix}
   =
   \mathcal{M}
   \begin{bmatrix}
     S_x(n)  \\
     S_y(n)  \\
     S_z(n)
   \end{bmatrix}
   =
   \begin{pmatrix}
     \cos\Xi_n & -\cos\alpha\sin\Xi_n  &  \sin\alpha\sin\Xi_n  \\
     \sin\Xi_n  & \cos\alpha\cos\Xi_n  &  -\sin\alpha\cos\Xi_n \\
             0              &        \sin\alpha                    &            \cos\alpha             
   \end{pmatrix}
   \begin{bmatrix}
     S_x(n)   \\
     S_y(n)   \\
     S_z(n)
   \end{bmatrix},
 \ee 
 where $\Xi_n=\kappa[S_y(n)\sin\alpha+S_z(n)\cos\alpha]$.
 As the classical angular momentum ${\bf S}=(S_x, S_y, S_z)$ is unit vector, it can be parametrized in terms of 
 the azimuthal angle $\theta$ and polar angle $\phi$ as ${\bf S}=(\cos\phi\sin\theta,\sin\phi\sin\theta,\cos\theta)$.
 Hence, the classical phase space is a two dimensional space with variables $\phi=\arctan(S_y/S_x)$ and 
 $\theta=\arccos(S_z)$.
 
\begin{figure}
  \includegraphics[width=\columnwidth]{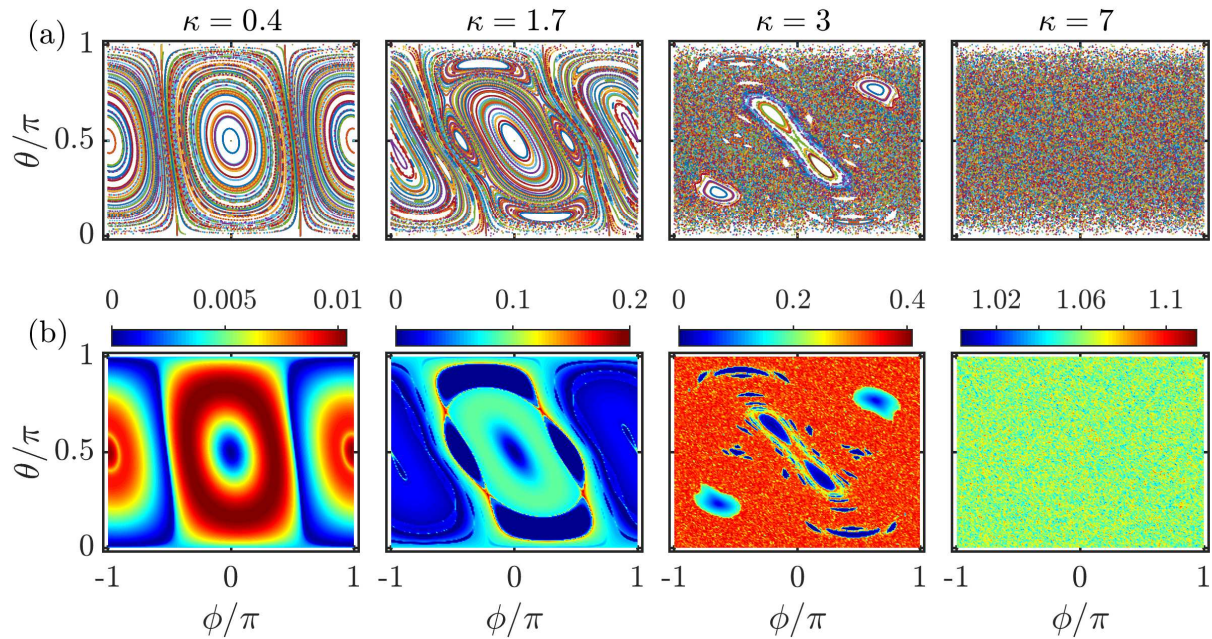}
  \caption{Row (a): Phase-space portraits of the classical kicked top.
  The classical variables $(\phi,\theta)$ are plotted for 289 random initial 
  conditions, each evolved for $300$ kicks. 
  Row (b): Color scaled plots of the largest Lyapunov exponent of the classical kicked top 
  for different initial conditions.
  The largest Lyapunov exponents are calculated on a grid with $200\times200$ initial conditions, 
  each evolved for $5000$ kicks.
  The different columns correspond to (from left to right): $\kappa=0.4, 1.7, 3$ and $\kappa=7$.
  Other parameter: $\alpha=4\pi/7$. 
  All quantities are dimensionless.}
  \label{CpsandLyE}
 \end{figure}

 It is known that the classical kicked top model is integrable at $\kappa=0$ and shows 
 increasingly chaotic behavior with increasing $\kappa$.
 To visualize how the value of $\kappa$ affects the dynamics of the classical kicked top model, 
 the phase-space portraits for different $\kappa$ values with $\alpha=4\pi/7$ are plotted in Fig.~\ref{CpsandLyE}(a). 
 The phase space is largely dominated by the regular orbits at small values of $\kappa$, 
 as shown in the first two columns of Fig.~\ref{CpsandLyE}(a).
 The phase space becomes mixed with regular regions coexisting with the chaotic sea as $\kappa$ is increased, see 
 the third column of Fig.~\ref{CpsandLyE}(a).
 For $\kappa$ increaseing further, the phase space is fully covered by chaotic trajectories, 
 there is no visible regular island in the last column of Fig.~\ref{CpsandLyE}(a).

 To quantify the chaotic features observed in Fig.~\ref{CpsandLyE}(a),  
 we investigate the behavior of the largest Lyapunov exponent of the classical map in Eq.~(\ref{CMP}).
 The largest Lyapunov exponent measures the rate of divergence between two infinitesimally close orbits of 
 a dynamical system \cite{Caiani1997,Vallejos2002,Arias2021}. 
 The largest Lyapunov exponent, therefore, estimates the level of chaos.
 For the classical map in Eq.~(\ref{CMP}), the largest Lyapunov exponent is defined as \cite{Parker2012}
 \be
    \lambda_+=\lim_{n\to\infty}\frac{1}{n}\ln\left[\frac{||\delta\mathbf{S}(n)||}{||\delta\mathbf{S}(0)||}\right],
 \ee
 where the Oseledets ergodic theorem \cite{Oseledets1968} guarantees the existence of the limit.
 Here, the 3-dimensional vector $\delta\mathbf{S}(n)$ is the tangent vector associated with $\mathbf{S}(n)$ and 
 satisfies the following tangent map
 \be
     \delta\mathbf{S}(n+1)=\mathcal{T}[\mathbf{S}(n)]\delta\mathbf{S}(n)
        =\left[\frac{\partial\mathbf{S}(n+1)}{\partial\mathbf{S}(n)}\right]\delta\mathbf{S}(n),
 \ee
 with initial condition $\delta\mathbf{S}(0)$.
 Then, the largest Lyapunov exponent of the classical kicked top can be calculated as \cite{Arias2021,Constantoudis1997}
 \be
     \lambda_+=\ln\left[\lim_{n\to\infty}(\mu_+)^{1/n}\right],
 \ee 
 where $\mu_+$ denotes the largest eigenvalue of the matrix $\prod_{\ell=1}^n\mathcal{T}[\mathbf{S}(\ell)]$.
 In the limit of strong chaotic dynamics $\kappa\to\infty$ it has been found that the largest Lyapunov exponent 
 has the following approximate expression \cite{Constantoudis1997}
 \be\label{AppLy}
     \lambda_+^\infty=\ln(\kappa\sin\alpha)-1,
 \ee
 where $\sin\alpha>0$. 
 It has been shown that the classical map in Eq.~(\ref{CMP}) has no fully developed chaos for the cases of
 $\alpha=k\pi$ with $k=0,1,2,\ldots$ \cite{Bhosale2018}. 
 This is due to the fact that the angle $\theta$ 
 either keeps fixed at $\arccos[S_z(0)]$ or oscillates between $\arccos[S_z(0)]$ 
 and $\pi-\arccos[S_z(0)]$ in these cases. 
 On the other hand, the cases of $\alpha=(2k+1)\pi/2$ allow the strongest chaotic dynamics for classical kicked top.  
 
  \begin{figure}
    \includegraphics[width=\columnwidth]{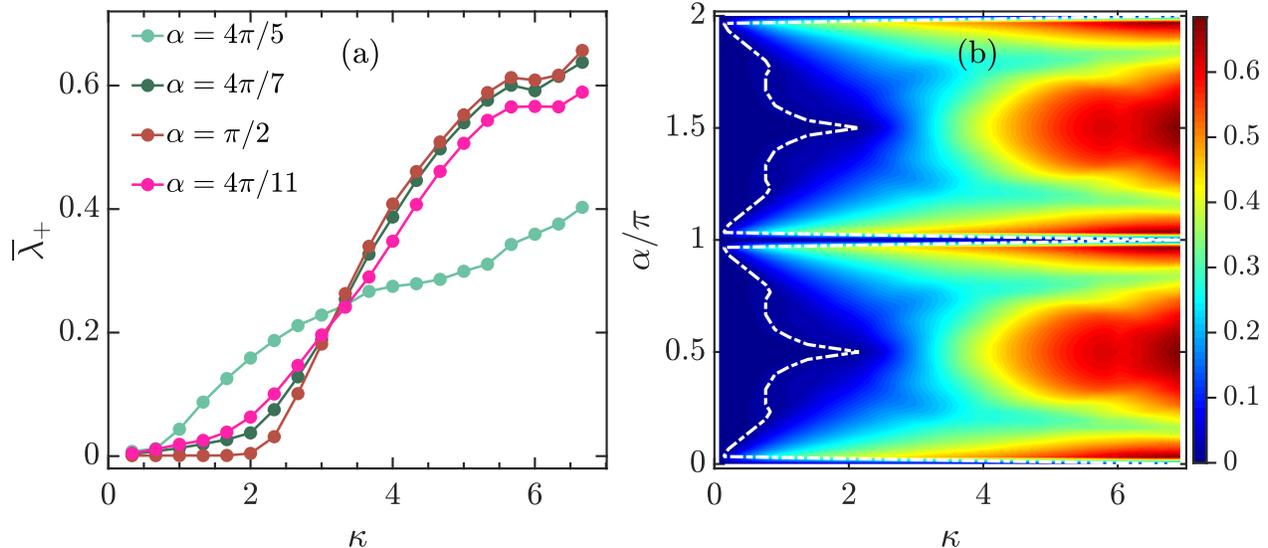}
  \caption{(a): Phase space averaged largest Lyapunov exponent $\bar{\lambda}_+$ as a function of $\kappa$ for
  several values of $\alpha$.
  (b): $\bar{\lambda}_+$ as a function of $\kappa$ and $\alpha$.
  The averaged largest Lyapunov exponents are calculated by averaging $\lambda_+$ over $40000$ 
  different initial conditions, each evolved for $5000$ kicks.
  In (b) the white dot-dashed curve corresponds to the values of $\kappa_c$ at which 
  $\bar{\lambda}_+=0.002$. 
  All quantities are dimensionless.}
  \label{LyEkp}
 \end{figure}

 In the row (b) of Fig.~\ref{CpsandLyE}, the largest Lyapunov exponents for different initial 
 points in the $\phi-\theta$ plane corresponding to the same values of $\kappa$ used in row (a) are plotted.
 By comparing Figs.~\ref{CpsandLyE}(b) and \ref{CpsandLyE}(a), we found that the largest Lyapunov
 exponents demonstrated remarkable resemblance with the corresponding classical phase portraits. 
 The dominated regular orbits at small $\kappa$ in the phase space leads to the tiny values of the
 largest Lyapunov exponents, as seen in the first two columns of Fig.~\ref{CpsandLyE}(b). 
 However, the fully chaotic phase space at $\kappa=7$ is clearly manifested by larger values
 of the largest Lyapunov exponent, which shows a uniform distribution in the phase space 
 [see the last column in Fig.~\ref{CpsandLyE}(b)]. 
 In particular, the regular regions in the mixed phase space are identified by $\lambda_+=0$, as depicted in
 the third column of Fig.~\ref{CpsandLyE}(b).   
 
  \begin{figure}
    \includegraphics[width=\columnwidth]{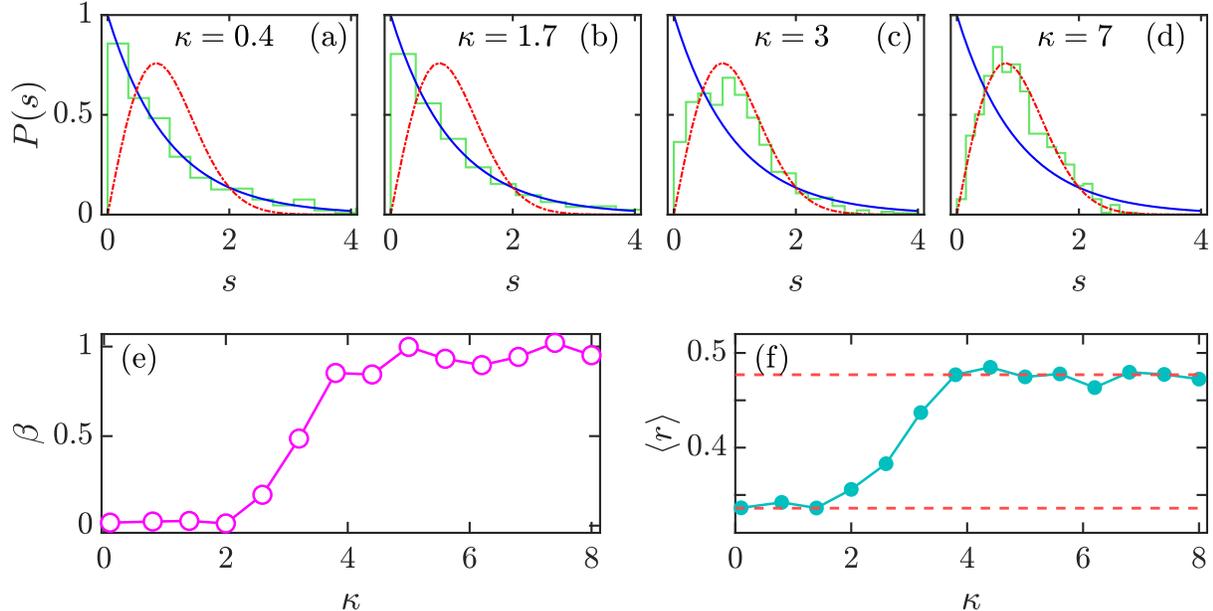}
  \caption{Level spacing distributions of the kicked top model for (a) $\kappa=0.4$, (b) $\kappa=1.7$, (c) $\kappa=3$,
  and (d) $\kappa=7$. The Poisson distribution is plotted as blue solid curve, while the red dot-dashed curve denotes 
  the Wigner-Dyson statistics.
  (e) The level repulsion exponent $\beta$ as a function of $\kappa$.
  (f) Averaged level spacing ratio $\la r\ra$ as a function of $\kappa$. The upper (bottom) red dashed line
  indicates $\la r\ra_{COE}\approx0.527$ ($\la r\ra_P\approx0.386$).  
  Other parameters: $j=1000$ and $\alpha=4\pi/7$.
  All quantities are dimensionless.}
  \label{Psbeta}
 \end{figure}

 To further reveal the effect of the kicking strength on the overall degree of chaos 
 in the classical kicked top, 
 we consider the phase space averaged largest Lyapunov exponent $\bar{\lambda}_+$, 
 which is defined as
 \be
    \bar{\lambda}_+=\frac{1}{4\pi}\int dS\lambda_+,
 \ee 
 where $dS=\sin\theta d\theta d\phi$ is the area element (or Haar measure) in the phase space \cite{Ariano1992}.
 It is interesting to note that $\bar{\lambda}_+$ can be considered as the rescaled 
 Kolmogorov-Sinai (KS) entropy $h_{KS}$ \cite{Kolmogorov1958,Kolmogorov1959},
 as according to the Pesin formula \cite{Pesin1977}, $h_{KS}$ of the kicked top model is equal to the 
 sum of the largest Lyapunov exponents, so that $h_{KS}=\int dS\lambda_+$.
   
 We plot $\bar{\lambda}_+$ as a function of $\kappa$ for different values of $\alpha$ in Fig.~\ref{LyEkp}(a).
 From this figure, we see that $\bar{\lambda}_+$ exhibits a rapid growth with increasing 
 $\kappa$ when $\kappa>\kappa_c$, regardless of the value of $\alpha$. 
 Here, $\kappa_c$ is defined as a threshold at which $\bar{\lambda}_+|_{\kappa=\kappa_c}=0.002$.
 This implies the onset of chaos in the classical kicked top for $\kappa>\kappa_c$. 
 We further observe that with change of $\alpha$ there is a variation in the value of $\kappa_c$. 
 Fig.~\ref{LyEkp}(b) depicts $\bar{\lambda}_+$ as a function of $\alpha$ and $\kappa$. 
 We make several observations from Fig.~\ref{LyEkp}(b).
 First, the behavior of $\bar{\lambda}_+$ shows a symmetry with respect to $\alpha=\pi$.
 This is because for the classical map in Eq.~(\ref{CMP}) $\alpha\to\alpha+\pi$ is equivalent to the transformation $S_x\to-S_x$,
 $S_y\to-S_y$ and $S_z\to-S_z$, which keeps the largest Lyapunov exponent unchanged \cite{Ariano1992}.
 Second, as the classical kicked top is integrable at $\alpha=0, \pi, 2\pi$, we have $\bar{\lambda}_+=0$  
 for these values of $\alpha$, regardless of $\kappa$.
 Finally, for $0<\alpha<\pi$, even though the sharp growth behavior of $\bar{\lambda}_+$ with increasing $\kappa$
 for $\kappa>\kappa_c$ is independent of $\alpha$, there is a strong dependence of 
 $\kappa_c$ on $\alpha$, as we have already seen in Fig.~\ref{LyEkp}(a).
 The white dot-dashed curve in Fig.~\ref{LyEkp}(b) shows how $\alpha$ affects the 
 value of $\kappa_c$.
 By confining to the range $0<\alpha<\pi$, we see that $\kappa_c$ is firstly increased with increasing $\alpha$ 
 and it reaches its maximal value at $\alpha=\pi/2$, then it starts to decrease as $\alpha$ increases further.
 The maximal value of $\kappa_c$ at $\alpha=\pi/2$ results from the additional symmetry of the system \cite{Haake1987},
 which leads to the onset of chaos occurring later than in the cases with other values of $\alpha$.
 Without loss of general qualitative behavior, in the following of this work, we fixed $\alpha=4\pi/7$.

  \begin{figure}
    \includegraphics[width=\columnwidth]{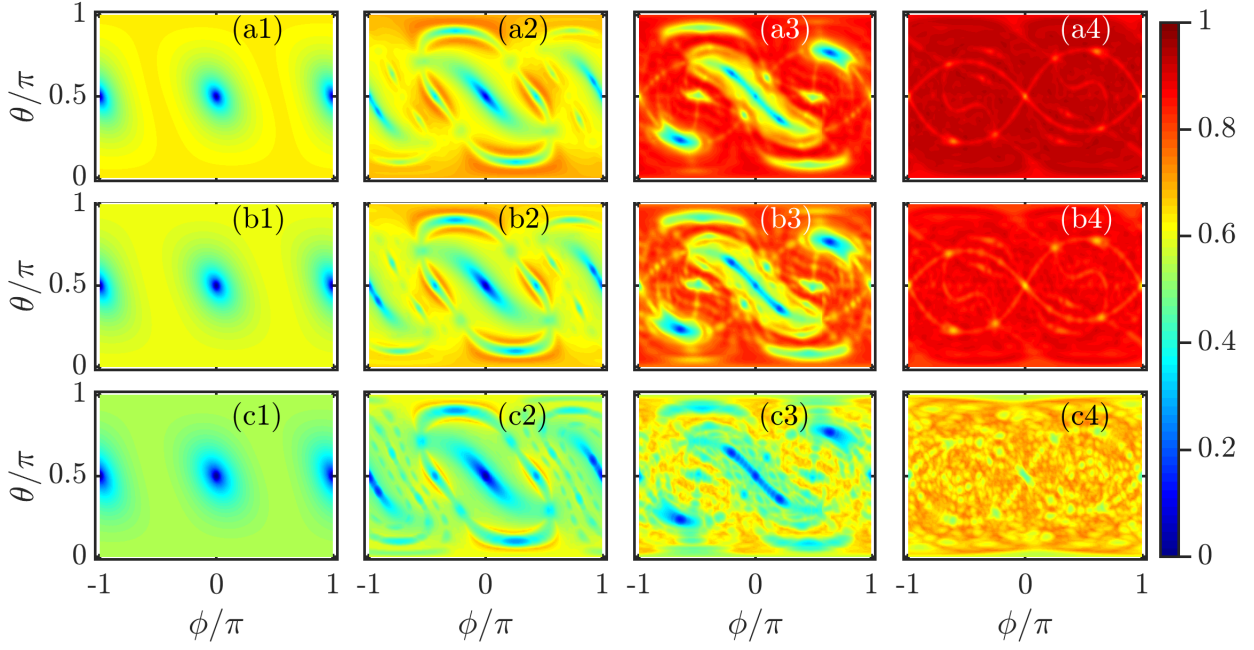}
  \caption{Color scaled plot of multifractal dimensions $D_q$ for (a1)-(a4) $q=1$, (b1)-b(4) $q=2$, 
  and (c1)-(c4) $q=\infty$, calculated on a grid of $100\times100$ coherent states.
  The different columns correspond to (from left to right): 
  $\kappa=0.4$, $\kappa=1.7$, $\kappa=3$, and $\kappa=7$. 
  Other parameters: $j=150$ and $\alpha=4\pi/7$.
  All quantities are dimensionless.}
  \label{Fd3d}
 \end{figure}

 \subsection{Quantum chaos of the kicked top model}
 
 The above discussed classical chaotic features are associated with quantum chaotic behavior in 
 quantum kicked top model. 
 The quantum character of chaos can be detected in several ways,
 such as the statistical properties of eigenvalues and eigenvectors 
 \cite{Haake2010,Izrailev1990,Brody1981}, 
 the dynamical features of entanglement entropy 
 \cite{WangX2004,Furuya1998,Arul2001,Seshadri2018,Gietka2019,Lerose2020},
 the decay in fidelity \cite{Emerson2002}, the correlation hole in survival probability \cite{Torres2017}, 
 and, in particular, the dynamics of the out-of-time-ordered correlator (OTOC)
 \cite{Rozenbaum2017,Mata2018,Carlos2019,Seshadri2018,Fortes2019,Lerose2020}.
 Among them, one of the most widely used is energy-level statistics of the quantum Hamiltonian.
 It is known that integrable systems allow level crossings which give rise to 
 Poisson distribution of the nearest level spacings \cite{Berry1977}. 
 On the other hand, based on the work of Wigner \cite{Wigner1958}, 
 Bohigas, Giannoni, and Schmit conjecture predicts that the energy-levels in chaotic systems should 
 exhibit level repulsion and the distribution of the nearest level spacings follows the 
 Wigner-Dyson distribution \cite{BGS1984}.
 {Here, we would like to point out that the explanation of the BGS conjecture
 has been first investgated through 2-point spectral correlation function \cite{Heusler2007,Haake2010}, 
 and then extended to $n$-point correlations with $n>2$
 \cite{Nagao2009,MullerS2018a,MullerS2018b}
 }.    
 
 The spectral statistics for a periodically driven quantum system can be analyzed through the 
 quasienergies (or eigenphases) of the Floquet {operator \cite{Zeldovich}}.
 The quasienergy spectrum of the kicked top model is obtained from the eigenphases of
 the Floquet operator $\mathcal{F}$ in Eq.~(\ref{Floquet}), and are defined as 
 \be
    \mathcal{F}|\nu_i\ra=e^{i\nu_i}|\nu_i\ra,
 \ee
 where $\nu_i$ denotes the $i$th eigenphase of $\mathcal{F}$ with corresponding eigenstate $|\nu_i\ra$.
 As $\{\nu_i\}$ are $2\pi$ periodic, we restrict them within the principal range $[-\pi,\pi)$. 
 
  \begin{figure}
    \includegraphics[width=\columnwidth]{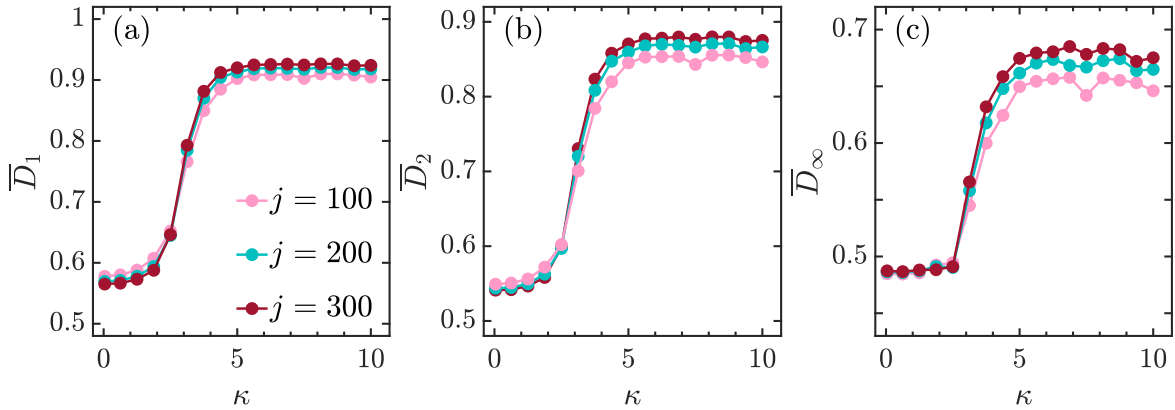}
  \caption{The variation of phase space averaged multifractal dimensions $\overline{D}_q$ 
  with kicking strength $\kappa$ for different $j$ are denoted by color scales.
  The phase space average is performed over $10^4$ coherent states in phase space. 
  Other parameters: $\alpha=4\pi/7$.
  All quantities are dimensionless.}
  \label{Frdkp}
 \end{figure}

 Numerically, the spectral analysis is performed as follows. 
 Firstly, we diagonalize $\mathcal{F}$ in the basis $\{|j,m\ra\}_{m=-j}^{m=j}$, and only consider 
 the quasienergies for the Floquet eigenstates with even parity. 
 Then, by arranging $\{\nu_i\}$ in ascending order, we define the gap between two consecutive
 levels as $d_i=\nu_{i+1}-\nu_i$. 
 Finally, we calculate the distribution $P(s)$ of the normalized level spacings $s_i=d_i/\la d\ra$
 \cite{Haake2010}, where $\la d\ra$ denotes the mean spacing. 
 The dependence of $P(s)$ on $\kappa$ is shown in Figs.~\ref{Psbeta}(a)-\ref{Psbeta}(d).
 Obviously, with increasing $\kappa$, the level spacing distribution $P(s)$ 
 undergoes a transition from Poisson statistics $P_P(s)=e^{-s}$ 
 to Wigner-Dyson statistics $P_{WD}(s)=(\pi/2)s\exp(-\pi s^2/4)$.
 This is consistent with the classical dynamics observed in Fig.~\ref{CpsandLyE}.
 
 To estimate the degree of chaos in Floquet spectrum of the kicked top model, we fit $P(s)$ to
 the so-called Brody distribution defined as \cite{Brody1981}
 \be
     P_B(s)=b_\beta(\beta+1)s^\beta\exp[-b_\beta s^{\beta+1}],
 \ee
 where the factor $b_\beta$ can be calculated as 
 \be
    b_\beta=\left[\Gamma\left(\frac{\beta+2}{\beta+1}\right)\right]^{\beta+1},
 \ee
 where $\Gamma(x)$ is the gamma function.
 The parameter $\beta$, which measures the degree of repulsion between levels, 
 is the level repulsion exponent
 and varies in the range $0\leq\beta\leq1$.
 For $\beta=0$, the Brody distribution reduces to Poisson distribution, 
 while it becomes Wigner-Dyson distribution
 at $\beta=1$. 
 Therefore, the larger $\beta$ is, the stronger the chaotic spectrum is.
 Fig.~\ref{Psbeta}(e) plots $\beta$ as a function of $\kappa$ with $j=1000$ and $\alpha=4\pi/7$.
 The behavior of $\beta$ nicely agrees with spectral analysis: 
 For $\kappa\lesssim2$, we have $\beta\approx0$, implying  
 the Poisson distribution of $P(s)$, while $\beta$ approaches unity when $\kappa\gtrsim5$, 
 suggesting that the quasienergy levels have the strongest repulsion and $P(s)$ being
 the Wigner-Dyson distribution. 
 It is worth pointing out that the transition region defined as $0<\beta<1$ 
 corresponds to the classical mixed phase space with regular regions embedded in the chaotic sea. 
 (see, e.~g. the third column in Fig.~\ref{CpsandLyE}). 
 More details about the spectral statistics in the transition region between integrability and chaos can be
 found in \cite{Prosen1994} and references therein. 
 We only mention that here the Berry-Robnik level spacing distribution 
 \cite{Berry1984} is not yet manifested 
 as we are not yet in sufficiently deep semiclassical regime and observe Brody distribution instead.
 
  \begin{figure}
    \includegraphics[width=\columnwidth]{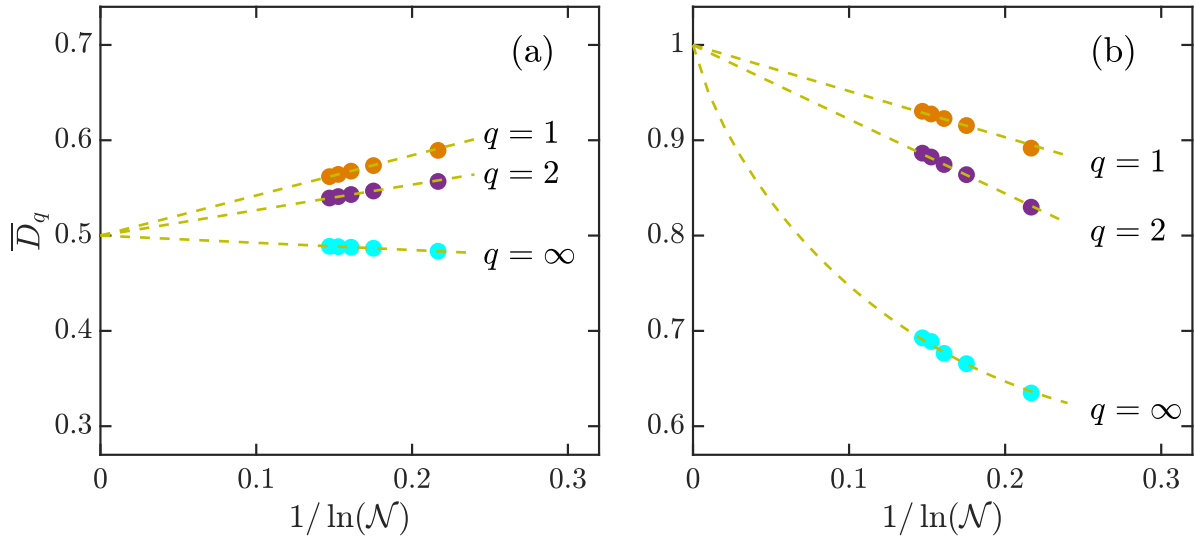}
  \caption{Phase space averaged fractal dimensions $\overline{D}_q$ 
  with $q=1,2,\infty$ versus $1/\ln\mathcal{N}$ 
  for $\kappa=0.4$ (a) and $\kappa=7$ (b).
  Here $\mathcal{N}$ denotes the dimension of Hilbert space.
  $\overline{D}_q$ were calculated from $10^4$ coherent states in phase space.
  Dashed lines in panel (a) are of the form $1/2-f_q/\ln\mathcal{N}$ 
  with $f_1=0.421, f_2=0.267$ and $f_\infty=-0.0758$.
  In panel (b), dashed lines for $q=1,2$ are of the form $1-g_q/\ln\mathcal{N}$ 
  with $g_1=0.484$ and $g_2=0.779$, 
  while the dashed line for $q=\infty$ is given by $1-g_\infty\ln(\ln\mathcal{N})/\ln\mathcal{N}$ 
  with $g_\infty=1.097$.
  Other parameters: $\alpha=4\pi/7$.
  All quantities are dimensionless.}
  \label{DqN}
 \end{figure}

 Besides the level spacing distribution, the mean ratio of consecutive level spacing is another widely 
 used detector of quantum chaos.
 Given the level spacing $\{d_i\}$, the mean ratio of level spacing is defined as \cite{Oganesyan2007,Atas2013}
 \be
    \la r\ra=\frac{1}{\mathcal{N}}\sum_{i=1}^{\mathcal{N}}r_i,\quad
    r_i=\mathrm{min}\left(\frac{1}{\delta_i},\delta_i\right),
 \ee
 where $\mathcal{N}$ is the total number of $r_i$ and $\delta_i=d_{i+1}/d_i$.
 It has been demonstrated that the averaged ratio of level spacing, $\la r\ra$, 
 acts as an indicator of spectral statistics.
 For regular systems with Poisson statistics $\la r\ra_P\approx0.386$ while $\la r\ra_{COE}\approx0.527$ for 
 circular orthogonal ensemble (COE) of random matrices \cite{Atas2013}.
 We plot $\la r\ra$ as a function of $\kappa$ for $\alpha=4\pi/7$ in Fig.~\ref{Psbeta}(f).
 One can see that $\la r\ra$ exhibits a crossover from 
 $\la r\ra_P$ to $\la r\ra_{COE}$ with $\kappa$ increasing. 
 This is in agreement with the behavior of $P(s)$, 
 as observed in Figs.~\ref{Psbeta}(a)-\ref{Psbeta}(d).
 Moreover, we notice that the behavior of $\la r\ra$ is similar to the level 
 repulsion exponent $\beta$ [cf.~Fig.~\ref{Psbeta}(e)].

 Even though the level statistics becomes a standard probe in the studies of quantum chaos, 
 it can not detect the local chaotic features in quantum systems.  
 In order to characterize the phase space structure and get more insights into the quantum-classical correspondence, 
 we consider the multifractal properties of the coherent states in the following.
 
 \subsection{Coherent states}
 
 The coherent states have wide applications in many fields 
 \cite{Klauder1985,ZhangW1990,Gazeau2009,Robert2012,Antoine2018}.
 As the uncertainty of coherent states tends to zero in the classical limit, 
 one can expect that the phase space structure and the quantum-classical 
 correspondence can be unveiled through appropriate properties of coherent states. 
 For our purpose, we use the generalized $\mathrm{SU}(2)$ coherent spin states, which are constructed by applying 
 an appropriate rotation on the state $|j,j\ra$ \cite{ZhangW1990,Gazeau2009},
 \be
    |\vartheta,\varphi\ra=\exp{\left[i\vartheta(J_x\sin\varphi-J_y\cos\varphi)\right]}|j,j\ra,
 \ee
 where $\vartheta, \varphi$ provide the orientation of $\mathbf{J}$.
 Further simplification of $|\vartheta, \varphi\ra$ is available by performing Taylor expansion and 
 the final result is given by \cite{Radcliffe1971, Gazeau2009} 
 \be \label{CSS}
     |\vartheta, \varphi\ra=\frac{e^{\zeta J_-}}{(1+|\zeta|^2)^{j}}|j,j\ra
          =\sum_{m=-j}^j\frac{\zeta^{j-m}}{(1+|\zeta|^2)^{j}}\sqrt{\frac{(2j)!}{(j+m)!(j-m)!}}|j,m\ra,
 \ee
 where $J_-=J_x-iJ_y$ and $\zeta=\tan(\vartheta/2)e^{i\varphi}$.
 It is straightforward to show that the uncertainty of the coherent spin state $|\vartheta,\varphi\ra$
 in Eq.~(\ref{CSS}) vanishes as $j\to\infty$.
 
 {Here, it is worth noting that the coherent states have been
 exploited to explore the quantum and classical structures of the kicked-top model 
 in several works \cite{Ariano1992,Kus1991}. 
 The  quantum-classical correspondence for various structures was established.
 In particular, those works have shown that some valuable information 
 of the scarred eigenstates, which are
 localized along the classical unstable periodic orbits, can be extracted from 
 the properties of the coherent states. 
 }
 
  \begin{figure}
  \includegraphics[width=\columnwidth]{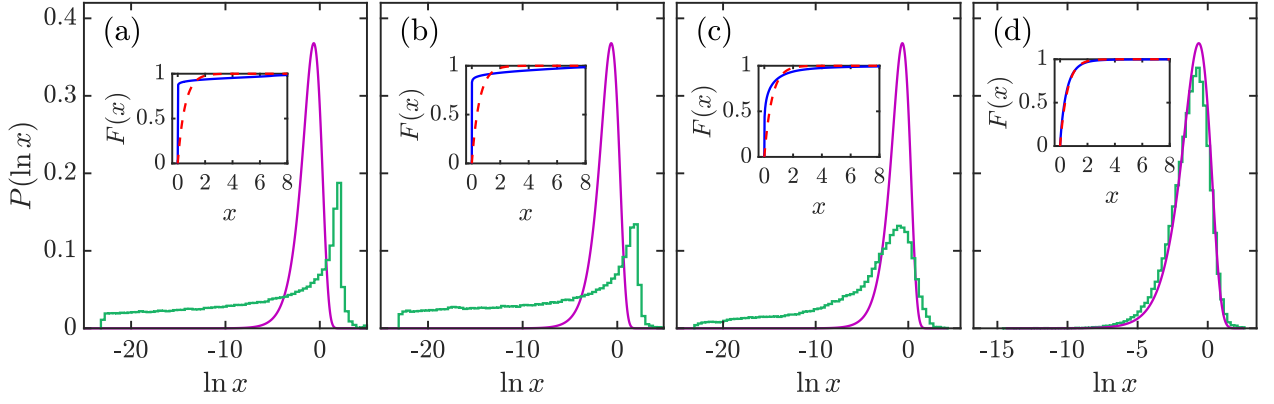}
  \caption{Histograms of $P(\ln x)$ for $\kappa=0.4$ (a), $\kappa=1.7$ (b), 
  $\kappa=3$ (c), and $\kappa=7$ (d).
  The purple solid lines in the main panels denote $P_2(\ln x)$ [cf.~Eq.~(\ref{Pdflog})].
  The inset in each panel plots their cumulative distributions 
  with blue solid curve corresponds to numerical result, while
  the red dashed curve represents $F_2(x)$ [cf.~Eq.~(\ref{CdFv})]. 
  $P(\ln x)$ has been computed from $10^4$ coherent states in phase space. 
  Other parameters: $j=150$ and $\alpha=4\pi/7$.
  All quantities are dimensionless.}
  \label{Prbx}
 \end{figure}

 \section{Multifractality of coherent states} \label{Third}
 
 The notion of multifractality was originally introduced to describe
 complex fluctuations observed in fluid turbulence \cite{Sreenivasan1991}.
 It has been recognized as a valuable tool to analyze a variety of classical complex phenomena. 
 Moreover, it has been found that the multifractal phenomenon was also visible in quantum state.
 Quantum state multifractality reflects its unusual statistical properties and has 
 attracted much attention as 
 it plays a prominent role in the phase transitions of different quantum systems 
 \cite{Ever2008,Stephan2011,Rodriguez2011,AtasY2013,Luitz2014,Lindinger2019,Mace2019,Pausch2021,Solorzano2021}.
 The characterization of the multifractality is quantified by the so-called generalized 
 fractal dimensions, denoted by $D_q$.
 To define $D_q$, let us consider a quantum state $|\Phi\ra$ expanded in a given orthonormal basis 
 $\{|k\ra\}$
 with dimension $\mathcal{N}$,
 \be
        |\Phi\ra=\sum_{k=1}^{\mathcal{N}}c_k|k\ra,
 \ee
 where $c_k=\la k|\Phi\ra$ and satisfies $\sum_k|c_k|^2=1$. 
 Then, $D_q$ is defined as \cite{Mirlin2000,Pausch2021}
 \be\label{FDq}
     D_q=\frac{S_q}{\ln\mathcal{N}}\quad  \text{and}\quad  S_q=\frac{1}{1-q}\ln\left(\sum_{k=1}^\mathcal{N}|c_k|^{2q}\right),
 \ee 
 where $S_q$ is the R\'{e}nyi entropy (or participation entropy).
 For finite $\mathcal{N}$, the values of $D_q$ are defined in the interval $D_q\in[0,1]$ and 
 decrease with increasing $q$ for $q\geq0$ \cite{Backer2019}.
 The fractal dimensions, $\mathcal{D}_q^\infty$, are obtained as $\mathcal{N}\to\infty$, so that
 $\mathcal{D}_q^\infty=\lim_{\mathcal{N}\to\infty}D_q$ \cite{Mirlin2000,Backer2019}.
 The degree of ergodicity of a quantum state in Hilbert space is measured by the fractal dimensions. 
 For a perfectly localized state $\mathcal{D}_q^\infty=0$ for $q>0$, whereas $\mathcal{D}_q^\infty=1 (\forall q)$ 
 corresponds to an ergodic state.  
 The multifractal states are the extended non-ergodic states and identified by $0<\mathcal{D}_q^\infty<1$.
 
  \begin{figure}
    \includegraphics[width=\columnwidth]{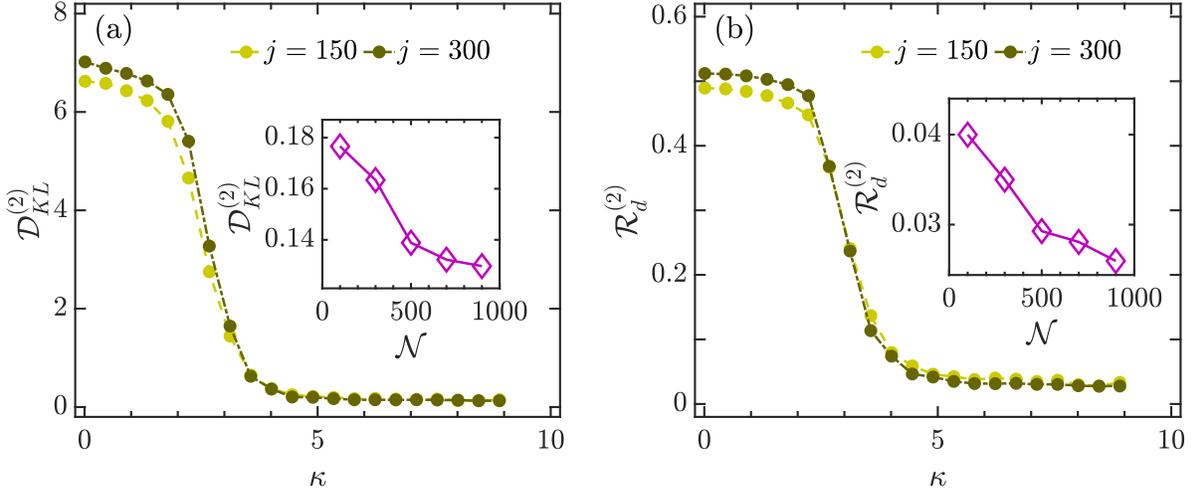}
  \caption{Panel (a): $\mathcal{D}^{(2)}_{KL}$ as a function of $\kappa$ for different system sizes.
  Inset: $\mathcal{D}^{(2)}_{KL}$ as a function of Hilbert space dimension $\mathcal{N}$ with $\kappa=8$.
  Panel (b): $\mathcal{R}^{(2)}_d$ as a function of $\kappa$ for different $j$ values.
  Inset: $\mathcal{R}^{(2)}_d$ versus Hilbert space dimension $\mathcal{N}$ for $\kappa=8$.
  Other parameter: $\alpha=4\pi/7$.
  All quantities are dimensionless.}
  \label{DRp}
 \end{figure}

 Among all $D_q$, we focus on the cases $q=1,2$ and $\infty$.
 As the R\'{e}nyi entropy reduces to the Shannon entropy, $S_1=-\sum_k|c_k|^2\ln|c_k|^2$, in the limit $q\to1$,
 the dimension $D_1$, also known as information dimension, controls the scaling of Shannon information entropy.
 For $q=2$, $S_2=\ln(\sum_k|c_k|^4)^{-1}$ is the logarithmic of the well-known participation ratio 
 \cite{Visscher1972,Mace2019,Solorzano2021}, 
 which measures the degree of delocalization of the state in Hilbert space.
 Hence, the exponent $D_2$ quantifies the scaling of the participation ratio.
 At $q=\infty$, the R\'{e}yni entropy turns into $S_\infty=-\ln p_m$ with $p_m=\mathrm{max}_k|c_k|^2$ and
 $D_\infty=-\ln p_m/\ln\mathcal{N}$, determining the extreme value statistics of the intensities of the quantum state.  
 
 In our study, we analyze the multifractal properties of the generalized  
 $\mathrm{SU}(2)$ coherent spin states [cf.~ Eq.~(\ref{CSS})] in the eigenvectors of the Floquet operator. 
 Therefore, we first expand $|\vartheta,\varphi\ra$ in the basis $\{|\nu_i\ra\}$ as follows
 \be
    |\vartheta,\varphi\ra=\sum_iw_i|\nu_i\ra,
 \ee
 where $w_i=\la\nu_i|\vartheta,\varphi\ra$ is the overlap between the basis vector $|\nu_i\ra$ and the coherent state
 $|\vartheta,\varphi\ra$, fulfilling the normalization condition $\sum_i|w_i|^2=1$.
 Then, by using Eq.~(\ref{FDq}), the fractal dimensions are calculated for coherent states that 
 are centered at different points $(\theta,\phi)$ of the classical phase space.
 
 In Fig.~\ref{Fd3d}, we plot $D_1, D_2$ and $D_\infty$ as a function of $\phi$ and $\theta$ 
 for different kicking strengths $\kappa$.
 By comparing with the classical phase space portraits in Fig.~\ref{CpsandLyE}(a), we observe that
 the underlying classical dynamics has strong effects on the properties of the fractal dimensions.
 The regular regions around the fixed points give rise to $D_q\approx0$, 
 indicating the coherent states located at these points are the localized states, as seen in the first and second
 columns of Fig.~\ref{Fd3d}.
 In the chaotic phase space, the fractal dimensions have larger values and exhibit 
 an approximately uniform distribution in the phase space (see the last column of Fig.~\ref{Fd3d}).
 These features imply that the coherent states have high degree of ergodicity for large kicking strength.
 For the mixed phase space, it is evident from the third column of Fig.~\ref{Fd3d} that
 the regular regions are identified by smaller fractal dimensions, while larger $D_q$ 
 correspond to the chaotic sea.
 The obvious correspondence between the fractal dimension and the classical phase space dynamics and the 
 Lyapunov exponents, as shown in Figs.~\ref{CpsandLyE}(a) and \ref{CpsandLyE}(b), suggests that
 $D_q$ are particularly useful to detect the signatures of quantum chaos.  
 We further notice that $0<D_q<1$ still holds even if the system is
 governed by regular dynamics.   
 
 To further demonstrate that $D_q$ can enable us to discern the regular and chaotic characters of the quantum system, 
 we assess the phase space averaged fractal dimensions, defined as
 \be
       \overline{D}_q=\frac{1}{4\pi}\int dS D_q.
 \ee
 Figs.~\ref{Frdkp}(a)-\ref{Frdkp}(c) show, respectively, $\overline{D}_1, \overline{D}_2$ and $\overline{D}_1$ as a function
 of $\kappa$ for different system sizes $j$.
 We see that the dependence of fractal dimensions on $\kappa$ are similar for different $j$. 
 The fractal dimensions change slowly with increasing $\kappa$ for smaller $\kappa$ and exhibit a rapid growth as
 soon as $\kappa>2$. Then $\overline{D}_q$ eventually approach their saturation values when $\kappa>5$.
 Moreover, we also observe that $\overline{D}_q$ are almost independent of $j$ for $\kappa<2$, while they increase with 
 increasing $j$ as long as $\kappa>5$. 
 
 In Fig.~\ref{DqN}, we plot the scaling of $\overline{D}_q$ with $1/\ln\mathcal{N}$ for $\kappa=0.4$ and $\kappa=7$.
 Here, $\mathcal{N}$ being the Hilbert space dimension of the system. 
 For the regular regime with $\kappa=0.4$ [Fig.~\ref{DqN}(a)], $\overline{D}_q$ follow the linear scaling of the form
 $\overline{D}_q=1/2-f_q/\ln\mathcal{N}$ with $f_q$ depending on the value of $q$.
 In particular, the scaling behaviors of $\overline{D}_q$ imply that 
 $\overline{D}_q$ tend to $1/2$ rather than zero  as $\mathcal{N}\to\infty$.
 On the other side, according to RMT, $\overline{D}_q$ in fully chaotic regime 
 obey the following asymptotic behavior \cite{Backer2019,Pausch2021,Fyodorov2015}
 \be\label{AspDq}
    \overline{D}^s_q=
    \begin{dcases*}
    1-\frac{g_q}{\ln\mathcal{N}} & for $q=1,2$,  \\  
    1-g_\infty\frac{\ln(\ln\mathcal{N})}{\ln\mathcal{N}} & for $q=\infty$.
   \end{dcases*}
 \ee
 As the kicked top becomes a strongly chaotic system at larger $\kappa$, one can expect that the 
 scaling behaviors of $\overline{D}_q$ should be in agreement with above results and should approach unity in 
 the thermodynamic limit. 
 This is indeed what we see in panel (b) of Fig.~\ref{DqN}, 
 which shows how $\overline{D}_q$ vary with $1/\ln\mathcal{N}$ at $\kappa=7$. 
 A good agreement between the numerical data and $\overline{D}_q^s$ in Eq.~(\ref{AspDq}) leads us to conclude 
 that the coherent states in strongly chaotic regime become ergodic in the eigenstates of the Floquet operator.
 
 More insights about the ergodic property of the coherent states in the eigenvectors of the Floquet operator can
 be obtained through the statistics of the rescaled expansion coefficients $\{x_i=\mathcal{N}|w_i|^2\}$.  
 For fully chaotic systems, it has been demonstrated that the 
 probability distribution of $\{x_i\}$ for different ensembles are unified in the $\chi_\nu^2$ distribution, 
 as known in Refs
 \cite{Mehta2004,HaakeZ1990,Zyczkowski1990,Alhassid1986,Alhassid1989},
 \be\label{PDF}
      P_\nu(x)=\left(\frac{\nu}{2\la x\ra}\right)^{\nu/2}\frac{x^{\nu/2-1}}{\Gamma(\nu/2)}\exp\left(-\frac{\nu x}{2\la x\ra}\right),
 \ee
 where $\la x\ra$ is the mean value of $\{x_i\}$ and $\nu=1,2,4$ for 
 orthogonal, unitary, and symplectic ensembles, respectively.
 In particular, $P_\nu(x)$ turns into the well known Porter-Thomas distribution \cite{Porter1956} when $\nu=1$.
 The width of the distribution becomes narrower with increasing $\nu$, indicating the larger of the value of $\nu$, the 
 smaller the fluctuations of $\{x_i\}$.
 
 For the coherent state considered here, the expansion coefficients are the complex numbers and their distribution 
 in the fully chaotic regime should be expected to be given by $\chi_\nu^2$ distribution with $\nu=2$
 \cite{Brickmann1987,Kus1988}.  
 Moreover, due to the large amount of small coefficients, we explore the distribution of $\{\ln x_i\}$ instead of $\{x_i\}$.
 From Eq.~(\ref{PDF}), $P_\nu(\ln x)$ is given by
 \be\label{Pdflog}
     P_\nu(\ln x)=\frac{P_\nu(x)}{d\ln x/dx}=\left(\frac{\nu}{2\la x\ra}\right)^{\nu/2}\frac{x^{\nu/2}}{\Gamma(\nu/2)}
                    \exp\left(-\frac{\nu x}{2\la x\ra}\right).
 \ee
 The relation $P_\nu(\ln x)=xP_\nu(x)$ implies that $P(\ln x)$ has the maximal value at $x=\la x\ra$. 
 
 In the main panels of Fig.~\ref{Prbx}, we show $P(\ln x)$ and compare with $P_2(\ln x)$ for several values of $\kappa$.
 The numerical data are obtained from $10^4$ coherent states that are uniformly located in the phase space.
 As expected, for the regular case with smaller $\kappa$ values, the larger number of small coefficients leads to 
 a larger fluctuations around its averaged value and a grater deviation from $P_2(\ln x)$,
 as shown in Figs~\ref{Prbx}(a) and \ref{Prbx}(b).
 However, the peak of $P(\ln x)$ around $\ln x\sim2$ results in moments of $P(x)$ are $q$-dependent, 
 which means non-zero multifractal dimensions $D_q (q>0)$ at smaller $\kappa$ values.
 With further increasing $\kappa$,
 the distribution of $P(\ln x)$ shifts its location to larger values of $\ln x$ and becomes narrower
 [Fig.~\ref{Prbx}(c)]. 
 For even larger $\kappa$ value, the distribution $P(\ln x)$ eventually converges to $P_2(\ln x)$, 
 as visible in \ref{Prbx}(d). 
 Here, we would like to point out that the peaks observed in panels (a) and (b) of Fig.~\ref{Prbx} have 
 nothing to do with the regularity and/or chaos. 
 In fact, the appearance of them depends on the computation basis that we used to expand the quantum state, 
 as has been stressed in Ref.~\cite{HaakeZ1990}.
 The regularity of a system is only manifested in the long flat tail of $P(\ln x)$.  
 
 The convergence between the distributions of $P(x)$ and $P_2(x)$ as $\kappa$ increases is 
 also confirmed in the behavior of the corresponding cumulative distributions.
 For the distribution $P(x)$, the cumulative distribution is defined as
 \be \label{CdF}
      F(x)=\int_0^x P(t)dt,
 \ee
 while the cumulative distribution of $P_\nu(x)$ is given by
 \be \label{CdFv}
    F_\nu(x)=\int_0^x P_\nu(t)dt=\dfrac{\gamma\left[\nu/2,\nu x/(2\la x\ra)\right]}{\Gamma(\nu/2)},
 \ee
 where $\gamma(s,x)=\int_0^x t^{s-1}e^{-t}dt$ is the lower incomplete gamma function.
 The insets in Fig.~\ref{Prbx} show $F(x)$ and $F_2(x)$ for different $\kappa$ values.
 It can be seen that the deviation between $F(x)$ and $F_2(x)$ decreases with increasing $\kappa$, 
 in accordance with the behavior of $P(\ln x)$ observed in main panels.
 
 To quantify the distance between $P(x)$ and $P_2(x)$, we use two different deviation measures, namely
 the square root of the Kullback-Leibler divergence (SKLD) \cite{Kullback1951,Wouter2018} 
 and the root-mean-square error (RMSE) \cite{Schervish2014,QianW2021}. 
 For the observed distribution $P(x)$ and predicted distribution $P_\nu(x)$,
 the SKLD (RMSE), denoted as $\mathcal{D}^{(\nu)}_{KL} (\mathcal{R}^{(\nu)}_d)$, measures the
 difference between observed and predicted probability (cumulative) distributions.  
 The definition of SKLD and RMSE are, respectively, given by
 \begin{align}
    &\mathcal{D}^{(\nu)}_{KL}=\left\{\int_{x_0}^{x_m}P(x)\ln\left[\frac{P(x)}{P_\nu(x)}\right]dx\right\}^{1/2}, \\
    &\mathcal{R}^{(\nu)}_d=\left\{\frac{1}{x_m-x_0}\int_{x_0}^{x_m}\left[F(x)-F_\nu(x)\right]dx\right\}^{1/2},
 \end{align}
 where $x_0$ and $x_m$ are the minimum and maximum values of $\{x_i\}$, respectively.
 Both $\mathcal{D}^{(\nu)}_{KL}$ and $\mathcal{R}^{(\nu)}_d$ are defined in the interval 
 $\mathcal{D}^{(\nu)}_{KL}, \mathcal{R}^{(\nu)}_d\in[0,\infty)$.
 When $\mathcal{D}^{(\nu)}_{KL}=\mathcal{R}^{(\nu)}_d=0$, we have $P(x)=P_\nu(x)$, 
 whereas larger $\mathcal{D}^{(\nu)}_{KL}, \mathcal{R}^{(\nu)}_d$ values imply 
 a larger deviation between $P(x)$ and $P_\nu(x)$.
 
 The variation of distance between $P(x)$ and $P_2(x)$, measured 
 by $\mathcal{D}^{(2)}_{KL}$ and $\mathcal{R}^{(2)}_d$, 
 with $\kappa$ for different $j$ values is shown in Fig.~\ref{DRp}.
 We see that $\mathcal{D}^{(2)}_{KL}$ and $\mathcal{R}^{(2)}_d$ behave in a similar way with increasing $\kappa$.  
 For the regular regime with weak kicking strength $\kappa<2$, both of them have high values and 
 decrease slowly as $\kappa$ is increased.
 This means that the coherent states are far from ergodicity in the regular regime.
 Then, they exhibit a rapid decrease in the region $2\lesssim\kappa\lesssim5$, which corresponds 
 to the crossover from the integrability to full chaos. 
 Finally, for $\kappa>5$, as the system becomes globally chaotic, 
 both $\mathcal{D}^{(2)}_{KL}$ and $\mathcal{R}^{(2)}_d$
 decrease to very small values and are almost independent of $\kappa$.
 Hence, the coherent states are ergodic states in fully chaotic regime.
 Moreover, the degree of ergodicity of coherent states in strongly chaotic regime 
 can be enhanced by increasing the system size, as illustrated in the insets of Fig.~\ref{DRp}.
 
{Here, an interesting point deserves discussing, namely the connection between 
 the fractal dimensions $D_q$ and
 other quantum chaos probes. Among all detectors of quantum chaos, we focus on spectral form 
 factor (SFF) and out-of-time-ordered correlators (OTOCs).  Both of them have been extensively
 used in numerous recent studies \cite{ChanA2018,Kos2018,Bertini2018,Friedman2019,Kobrin2021,Flack2020,PBraun2020,
 Khramtsov2021,Rozenbaum2017,Lewis2019,Rozenbaum2019, Mata2018,Carlos2019,Fortes2019,
 JLee2019,Bergamasco2019,Harrow2021}.}
 
 {Let us first consider the relation between $D_q$ and SFF.
 The SFF is a powerful tool for detecting the spectral properties of  a system and
 is definded as the Fourier transform of the two-point correlation function
 of the level density \cite{Leviandier1986}.
 It is known that the behavior of SFF for integrable systems is 
 drastically different from the chaotic systems, mainly due to the fact that the regular and
 chaotic systems have different spectral statistics \cite{Bertini2018}. 
 This means that the SFF can be used as an efficient and sensitive
 indicator of quantum chaos.
 As SFF measures the correlation between energy levels, while $D_q$ characterizes the complexity of
 a quantum states in a given basis, there is no obvious relation between them.
 Although, for some particular cases, $D_q$ and SFF have been connected in 
 several works \cite{Chalker1996,Bogomolny2011},  
 a more general connection between them is still an open question, beyond the scope
 of the present work. We will explore this subject in our future work.
 }
 
{We now discuss the comparison of $D_q$ with OTOCs.
 As the main criterion employed to decide whether a quantum system is chaotic or not, 
 OTOC quantifies the sensitivity with respect to the initial condition and 
 information scrambling in quantum systems. 
 It has been demostrated that both the early and late time behavior of OTOC serve as
 useful diagnostics of quantum chaos 
 \cite{Rozenbaum2017,Mata2018,Carlos2019,Fortes2019,Rammensee2018,Prakash2020}.
 Since the chaotic dynamics leads to a rapid growth and large long-time saturation value 
 in the behavior of OTOC, one can, therefore, expect that the growth rate of the OTOC 
 as well as its long-time saturation value may be correlated with $D_q$.
 However, a more detailed  and general connection between them
 remains an open question. 
 Till date, only a formal relationship between $D_2$ and OTOCs has been established
 \cite{Hosur2016,FanR2017}.
 }
 
 {We finally point out that the degree of extension of a quantum state usually
 increases with the degree of chaoticity of the system. 
 Hence, we believe that qualitatively similar results should be obtained 
 for generic quantum states and for other quantum systems.
 Moreover, our main conclusions still hold if the coherent states are expanded 
 in another more localized basis, even if the fractal dimensions are dependent on the choice
 of the basis.
 }

 \section{Conclusions} \label{Forth}
 
 In this work, we have explored the quantum characters of chaos 
 in the quantum kicked top model by means of multifractal analysis.
 The kicked top model is a prototype model in the studies of quantum chaos and 
 its experimental realization has been achieved in several experiments
 \cite{Chaudhury2009,Neill2016,Tomkovic2017,Meier2019}.
 The signatures of classical chaos have been revealed in various works. 
 It was known that the phase space of classical kicked top has complex structures during the 
 transition from regular to chaotic dynamics.
 Therefore, how to capture the local chaotic features in quantum system becomes a
 crucial point to understand the quantum-classical correspondence.
 Although the indicators of quantum chaos, such as level spacing statistics and mean ratio of level spacings,
 are able to unveil the global signatures of chaos in quantum systems, they cannot detect the local chaotic behaviors. 
 In the present work, with the help of the generalized coherent spin states, we have investigated the 
 local chaotic properties of quantum kicked top through the multifractal dimensions of coherent states.
 
 The multifractal analysis of the coherent states is performed by expanding them in orthonormal 
 basis composed by the eigenstates of the Floquet operator.
 We explicitly demonstrated that the regular regions in the mixed phase space 
 clearly correspond to small values of multifractal dimensions.
 For the strong chaotic case, the multifractal dimensions exhibit uniform distribution in phase space.  
 Moreover, we have shown that the phase space averaged multifractal dimensions serve as indicators of quantum chaos.  
 With kicking strength increasing, the averaged multifractal dimensions undergo a rapid growth, indicating the transition
 from regular to chaotic dynamics of the system.
 Coherent states within the strongly chaotic regime become ergodic with 
 multifractal dimensions tend to unity in the thermodynamic limit, in accordance with the predictions of RMT.      
 However, coherent states' multifractal dimensions in the regular regime are not equal to zero. 
 Instead, they approach a finite value as the system size goes to infinity. 
 
 To get more insights into the multifractal characters of the coherent states 
 and their connections with the underlying chaotic dynamics, we further investigated 
 the probability distribution of the expansion coefficients.
 Such distribution is expected to follow the so-called $\chi_\nu^2$ distribution for the fully chaotic systems.
 We have shown that the deviation between the distribution of coefficients and $\chi_2^2$ distribution 
 decreases as the kicking strength is increased. 
 For the kicking strengths that lead to the fully chaotic dynamics, the distribution of coefficients 
 exhibits a quite good agreement with $\chi_2^2$ distribution, implying the strong ergodicity of coherent states. 
 On the contrary, the distribution of coefficients in the regular regime displays a remarkable 
 difference from $\chi_2^2$ distribution and its long flat tail reveals the localization character of coherent states. 
 In particular, the non-zero fractal dimensions for the regular case can be understood 
 as a consequence of the sharp peak appearing in the probability distribution of coefficients. 
 As the existence of the peak in the distribution of coefficients for the regular regime 
 is a basis dependent phenomenon, one can therefore expect that 
 the fractal dimensions in regular systems should be equal to zero 
 if a suitable computation basis has been selected. 
 How to identify an appropriate basis used in the multifractality analysis is an 
 interesting topic for future studies.  
 We also discuss how to measure the distance between the observed distribution of coefficients 
 and the expected $\chi_2^2$ distribution.
 We have shown that the transition from regular to chaotic dynamics of the system can be
 identified by the dramatic decrease in the behavior of different distance measures.
 
 As a final remark, we would like to point out that the recent experimental advances 
 enable a direct observation of multifractality 
 of wave packets in several quantum systems \cite{Faez2009,Lemarie2010,Richardella2010,Sagi2012}.
 Hence, the multifractal properties of our studied Floquet system are readily accessible 
 to state-of-the-art experimental platforms.   
         
\acknowledgements
This research was funded by the Slovenian Research Agency (ARRS) under the grant 
number J1-9112. 
Q.~W. acknowledges support from the National Science Foundation of China 
under grant number 11805165,
Zhejiang Provincial Nature Science Foundation under grant number LY20A050001.

\bibliographystyle{apsrev4-1}
\bibliography{MfKT}

\end{document}